\documentclass[conference]{IEEEtran}
\usepackage{amsfonts}
\usepackage{amsmath}

\ifCLASSINFOpdf
  \usepackage[pdftex]{graphicx}
  \graphicspath{{../pdf/}{../jpeg/}{Images/}}
  \DeclareGraphicsExtensions{.pdf,.jpeg,.png}
\else
\fi
\IEEEoverridecommandlockouts

\usepackage{tikz}
\usepackage{textcomp}
\usepackage{hyperref}
\usepackage{lipsum}

\newcommand\copyrighttext{
  \footnotesize \textcopyright 2017 IEEE. Personal use of this material is permitted.
  Permission from IEEE must be obtained for all other uses, in any current or future
  media, including reprinting/republishing this material for advertising or promotional
  purposes, creating new collective works, for resale or redistribution to servers or
  lists, or reuse of any copyrighted component of this work in other works.
  DOI: \href{https://doi.org/10.1109/ICTAI.2017.00151}{10.1109/ICTAI.2017.00151}}
\newcommand\copyrightnotice{
\begin{tikzpicture}[remember picture,overlay]
\node[anchor=south,yshift=10pt] at (current page.south) {\fbox{\parbox{\dimexpr\textwidth-\fboxsep-\fboxrule\relax}{\copyrighttext}}};
\end{tikzpicture}
}

\begin{document}
\title{Intelligent Fault Analysis in Electrical Power Grids}


%
\author{\IEEEauthorblockN{Biswarup Bhattacharya}
\IEEEauthorblockA{Department of Computer Science\\
University of Southern California\\
Los Angeles, CA 90089. USA.\\
Email: bbhattac@usc.edu}
\and
\IEEEauthorblockN{Abhishek Sinha}
\IEEEauthorblockA{Adobe Systems Incorporated\\
Noida, UP 201301. India.\\
Email: abhishek.sinha94@gmail.com}}

\maketitle
\copyrightnotice

\begin{abstract}
Power grids are one of the most important components of infrastructure in today's world. Every nation is dependent on the security and stability of its own power grid to provide electricity to the households and industries. A malfunction of even a small part of a power grid can cause loss of productivity, revenue and in some cases even life. Thus, it is imperative to design a system which can detect the health of the power grid and take protective measures accordingly even before a serious anomaly takes place. To achieve this objective, we have set out to create an artificially intelligent system which can analyze the grid information at any given time and determine the health of the grid through the usage of sophisticated formal models and novel machine learning techniques like recurrent neural networks. Our system simulates grid conditions including stimuli like faults, generator output fluctuations, load fluctuations using Siemens PSS/E software and this data is trained using various classifiers like SVM, LSTM and subsequently tested. The results are excellent with our methods giving very high accuracy for the data. This model can easily be scaled to handle larger and more complex grid architectures.
\end{abstract}
\IEEEpeerreviewmaketitle

\section{Introduction}
Power grids are now been considered to be one of the important components of infrastructure on which the modern society depends. The primary objective of power system operation is to supply uninterrupted power to the customers. But small and large scale faults and disturbances in the grid often cause power outages and thereby affect the system reliability and customer satisfaction.

Electrical power grids are huge systems, especially at the national level. The ``synchrophasors project'' in India \cite{synchrophasor} has been deployed with the help of which system operators are now able to monitor the magnitude and angle of each phase of the three phase voltage and current, frequency, rate of change of frequency and angular separation at every few millisecond intervals ($40$ milliseconds) at a Load Dispatch Centre (LDC). Thus the transient or dynamic behavior of the power system can be observed in near real-time at the control centre which hitherto was possible only in offline mode in the form of Substation Disturbance Records or through offline dynamic simulations performed on network models.

A phasor measurement unit (PMU) is a device which measures the electrical waves on an electricity grid using a common time source for synchronization. PMUs provide us with a huge amount of data \cite{pmu} and monitoring such huge data manually is an infeasible task. Thus, automated algorithms are required for monitoring the power grids, determining the health of the grid and performing intelligent fault analysis in real-time.

The importance of fault analysis lies in the fact that electrical faults cause maintenance issues, financial losses for companies and general inconvenience to consumers. Typically, electrical faults take multiple hours, if not days, to repair. Such kind of inefficient infrastructure causes a lot of problems to commercial establishments and households as almost all civilizations are nowadays heavily dependent on electricity to perform almost every task. Thus, with the help of intelligent fault analysis, one can, in a way, predict faults before they are just about to happen so as to ensure that the relays and the protective infrastructure in the power grid switch on in time and function as required. With accessibility to cutting edge technologies like recurrent neural networks, such kind of prediction capabilities are now possible to be performed within milliseconds and with high accuracy. Machine learning and deep learning models have been explored in mild detail previously in research literature, however we demonstrate with clarity that neural networks can indeed perform extremely well when applied in power system domains and especially electrical fault analysis. The approach we have used is such that it can even be scaled up to handle all grid sizes, from the small simulator grids to large real grids, like the Indian electricity grid, and can be deployed in the real world.

\section{Related Work}
Majority of the literature that exists on fault analysis is usually focused on numerical optimization techniques and statistical techniques. Relevant literature does exist for the kind of work, though sparingly, where some have applied artificial intelligence techniques on power systems to derive various relations. However, the methodology adopted by us using state-of-the-art techniques and achieving great accuracy has not been attempted before.

Some of the existing literature on fault analysis are discussed here. In \cite{related1}, a scheme was proposed which first identifies fault locations using an iterative estimation of load and fault current at each line section and then an actual location is identified, applying the current pattern matching rules. In \cite{related2}, expert system for the diagnosis of faults was discussed. In \cite{related4}, the paper proposes alternatives to improve the electric power service continuity using the learning algorithm for multivariable data analysis (LAMDA) classification technique to locate faults in power distribution systems. This was a data analytic approach to find location of faults whereas we are using a machine learning technique. With respect to AI approaches, in \cite{related3}, the authors discuss their experience of developing a multi-agent system that is robust enough for continual online use within the power industry. In \cite{related5}, the paper presents an artificial neural network (ANN) and support vector machine (SVM) approach for locating faults in radial distribution systems. Our approach is novel compared to this paper as we realized that long short-term memories (LSTMs) may be the better way to represent time series voltage and angle data and the hypothesis has been confirmed by our experiments.

\section{Background \& Preliminaries}

\subsection{Support Vector Machines (SVM)}
In machine learning, support vector machines (SVMs) are supervised learning models that analyze data and provide output for classification and regression analysis \cite{svm}. SVM is a non-probabilistic binary linear classifier, such that it classifies examples to be falling into one of two classes. The SVM margin, which is the separation between the two classes, is trained to be as optimal as possible. New examples are then mapped into that same training space and predicted to belong to one of the classes depending on which side of the hyperplane they fall in.

\subsection{Recurrent Neural Networks (RNN)}
The idea behind RNNs is to make use of sequential information. In a traditional neural network, it is assumed that all inputs and outputs are independent of each other. However, that may not be a useful choice of architecture always. For example, when trying to predict words that will come in a sentence, it is important to consider the words that came before, thereby incorporating a type of ``temporal'' or ``sequential'' aspect to the learning process. RNNs are called recurrent because they perform the same task for every element of a sequence, with the output being depended on the previous computations. RNNs have a ``memory'' which captures information about what has been calculated so far \cite{rnn}. A typical RNN structure has been shown in Figure 1.
\begin{figure}[h]
\centering
\includegraphics[width=\linewidth, keepaspectratio]{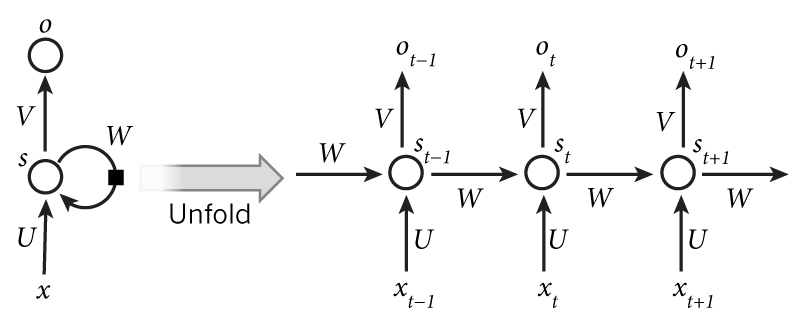}
\caption{A recurrent neural network and the unfolding in time (image from \cite{rnnpic}}
\end{figure}

Since in practice RNNs are limited in the span of time they can look back to, a variant of it, called as a LSTM (Long Short Term Memory) network, is used to take care of long time dependencies. Each LSTM cell gives as an output the hidden state till and the output vector at that point of time \cite{6hochreiter}. The LSTM cell structure is shown in Figure 2. 

\begin{figure}[h]
\centering
\includegraphics[width=\linewidth, keepaspectratio]{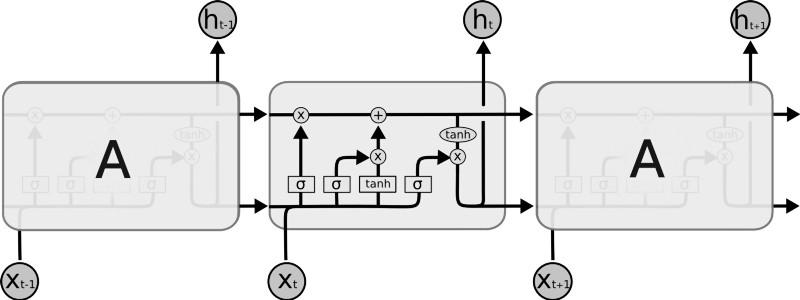}
\caption{Basic structure of a LSTM cell (image from \cite{lstmpic})}
\end{figure}

\section{Problem Description}
In the current situation of Indian electrical power grids, when a small disturbance is seen at a dispatch center, then the norm is to generate a report and check with other dispatch centers. If the disturbance is found to be local, then it is ignored. Else, if it is found to be correlated (similar disturbances observed at other dispatch centers), then further diagnostics are conducted. Our goal is to perform this fault analysis automatically using machine learning.

We aim to monitor the grid continuously and then in the case of a fault determine the type of the fault as and when it occurs. Apart from just predicting the nature of the fault we aim to give further insight into the fault such as in which bus line the fault had been triggered. The entire process has to be done automatically using machine learning techniques without using any human supervision and using just the current as well as past states of the grid as an input.

\section{Dataset}
\subsection{Software for Data Collection}
The experiments required the use of power grid and power system data. For our simulation and testing purposes, we used simulation data from the Siemens PSS/E software \cite{1siemens} and PowerWorld Simulator \cite{2powersim}.

The Siemens PSS/E software package can do fast and robust power flow solution for network models up to 200,000 buses. It has an array of features useful for dynamic and transient analysis. It also has an useful scripting system named {\tt{psspy}}. Using {\tt{psspy}}, one can use the modules of PSS/E remotely through Python. We used this scripting system for porting data, simulating and creating the dataset.

\subsection{Methodology}
\subsubsection{Inputs and Disturbances}
One can modify a huge number of parameters through PSS/E. However, for our analysis, we focused on mainly the power injection and load values at each bus. Keeping in mind our goal, it made sense to vary only the power values (real \& reactive) as that will correspond to a change in the bus voltages and angles. Using these bus voltages and angles, we can perform our analysis.

Using {\tt{psspy}}, we changed the power input and load at every bus at a certain timestamp suddenly during the simulation. This essentially mimics the fluctuations which happens in the real grids all the time. The value of the fluctuation was distributed along a uniform distribution with suitable upper and lower limits. Hence, the fluctuation could be very small in one bus whereas very large in another bus. We also made sure that no bus becomes totally disconnected in the grid due to too high fluctuations, as that essentially corresponds to a fault.

The network (electric grid) used by us consisted of $23$ buses with $6$ generators and $8$ loads at various buses throughout the network. The base frequency used was $50$ Hz which is the standard for India. The base MVA for the network was $100$ MVA.

\subsubsection{Outputs}
In the PMUs, one can measure the values of voltage and angle in real time. Hence, given the inputs and disturbances, we logged the voltage and angle at every bus at every timestamp. The logging was done at every $40$ ms which mimics the actual metering done by the PMUs. For every simulation, the data was collected for upto $4$ seconds of simulation time to capture the transient and also observe the changes due to generator or load fluctuations and faults suitably.

\subsubsection{Simulating the faults}
In an electric power system, a fault or fault current is any abnormal electric current \cite{4kothari}. For example, a short circuit is a fault in which current bypasses the normal load. An open-circuit fault occurs if a circuit is interrupted by some failure. In three-phase systems, a fault may involve one or more phases and ground, or may occur only between phases. In a ``ground fault'' or ``earth fault'', current flows into the earth. The prospective short circuit current of a predictable fault can be calculated for most situations. In power systems, protective devices can detect fault conditions and operate circuit breakers and other devices to limit the loss of service due to a failure.

In a polyphase system, a fault may affect all phases equally which is a ``symmetrical fault''. If only some phases are affected, the resulting ``asymmetrical fault'' becomes more complicated to analyze. The analysis of these types of faults is often simplified by using methods such as symmetrical components.

For our analysis, we focused on simulating the symmetrical faults and unsymmetrical faults. The faults simulated were namely:
\begin{itemize}
    \item $3\phi$ bus fault: This is a symmetrical fault as it affects all the phases at a bus equally.
    \item Branch Trip fault: This is a symmetrical fault which trips the transmission line (all $3$ phases) between two buses.
    \item LL fault: This is an unsymmetrical fault and it short circuits two phases (in PSS/E, these are Phase A and Phase B).
    \item LG fault: This is an unsymmetrical fault and it short circuits one phase (in PSS/E, this is Phase A) with the ground.
\end{itemize}

\begin{figure}[h]
\centering
\includegraphics[width=\linewidth, keepaspectratio]{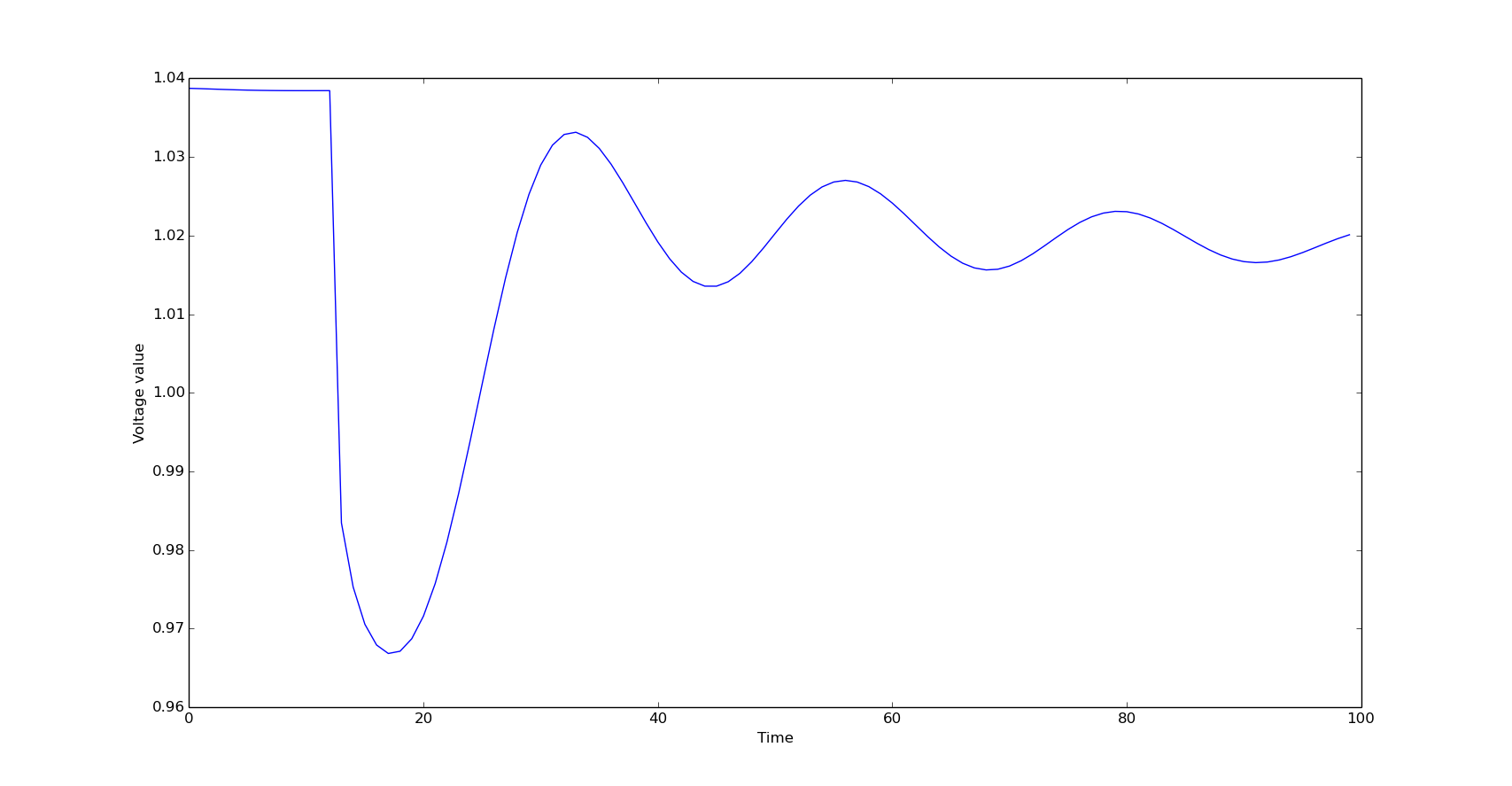}
\caption{A typical voltage varying voltage plot for a particular bus line in the presence of a fault}
\end{figure}

The voltage profile during a fault is shown in Figure 3. The fault was initiated and subsequently cleared at points during the simulation pre-decided by the data generator. This was however not known to the classifier built and the classifier was able to predict this time of initiation and clearing of faults without this knowledge simply by learning the data collected.

\subsection{Data Collected}
We collected data for all the $4$ fault types with generator or load fluctuations as specified. We simulated every type of fault when the fault occurs at each individual bus. Thus, this translates to data corresponding to $23$ buses for all $4$ types of faults. For each bus, the simulation was run $100$ times to generate different variations so that there are enough examples. Also, there were examples which consisted of just the generator or load fluctuations, and no faults so as to make the machine learn to differentiate between the allowed generator or load fluctuations and the fault conditions.

Each training example consisted of the volt and angle information for the particular bus at timestamps ranging from $0$ seconds to $4$ seconds with an interval of $0.04$ seconds. The amount of time required to generate the data on the server provided to us was around $20$ minutes per bus per fault ($100$ simulations per bus). 

\section{Forecasting the maximum voltage deviation using pre-fault data}
\subsection{The Model}
The data corresponding to a network having $23$ different buses and subjected to transmission fault (line trip) was obtained. A \textit{neural network model} \cite{5penman} was constructed to predict the maximum voltage deviations corresponding to each bus line just by looking at the pre-fault state of the network. The model consists of $2$ \textit{hidden layers} consisting of $60$ and $40$ neurons respectively. The input to the model is a vector of size equal to the number of bus lines corresponding to the pre-fault voltage data of each bus. Since the post fault voltage value will depend on the states of each of the buses, thus the prediction for each bus is not done independently, but rather the forecasting of entire network is done simultaneously. The output is of the same size as the input and contains the forecasted voltage value for each bus. The data was divided into $80$\% training and $20$\% test data and the results were obtained on test set after having trained the model on training data. The model used by us has been shown in Figure 4.

\begin{figure}[h]
\centering
\includegraphics[width=\linewidth, keepaspectratio]{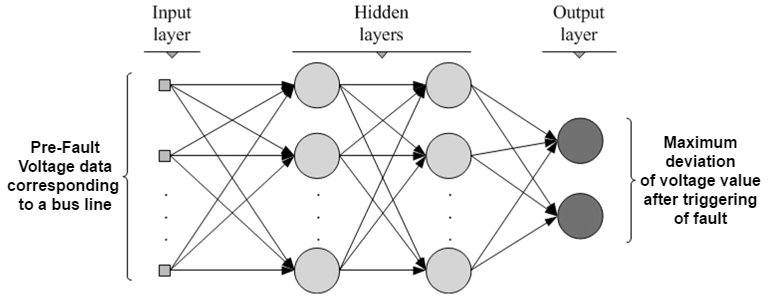}
\caption{Model for prediction of maximum voltage deviation after fault triggering (image adapted from \cite{2layernn})}
\end{figure}

\subsection{Results}
After $2000$ steps of training the following losses were obtained between the actual output and predicted output corresponding to the network. The mean $L_{2}$ error was equal to $2.8 \times 10^{-3}$ and the mean $L_{1}$ error was equal to $2.3 \times 10^{-2}$. The plot of $L_{2}$ error with progress of training has been shown in Figure 5.

\begin{figure}[h]
\centering
\includegraphics[width=\linewidth, keepaspectratio]{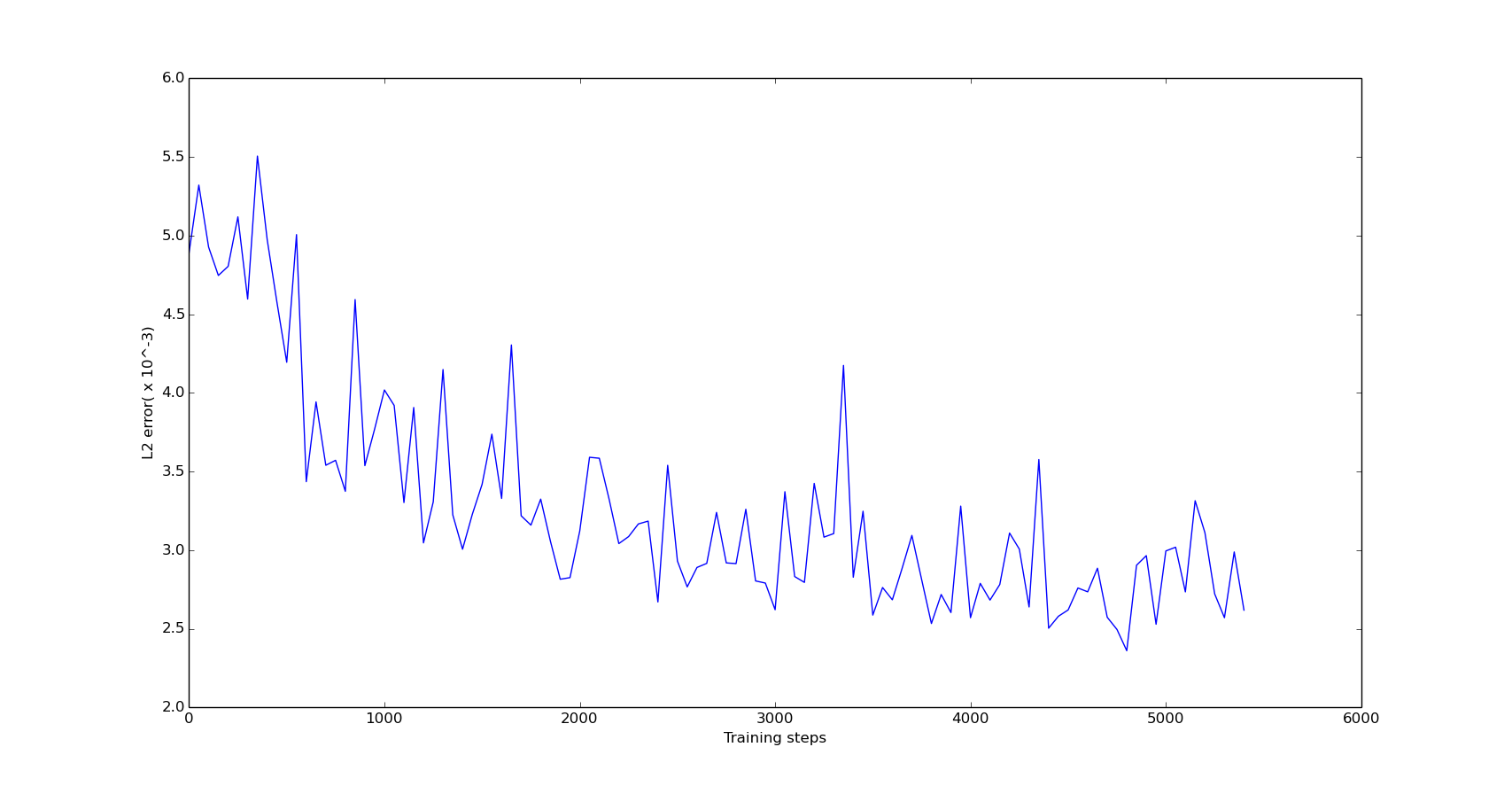}
\caption{Variation of $L_2$-error with progress of training}
\end{figure}

\section{Classification of faults}
Classification of faults given the entire network data was done for the case of \textit{LL (line to line)} and \textit{LG (line to ground)} faults. The voltage data corresponding to $100$ different time steps and for each of the bus lines present in the network are given as an input to the classifier. The classifier gives an output probability corresponding to the two fault classes. The variation of voltage values in the presence of LL and LG faults have been shown in Figures 6 and 7 respectively.

\begin{figure}[h]
\centering
\includegraphics[width=\linewidth, keepaspectratio]{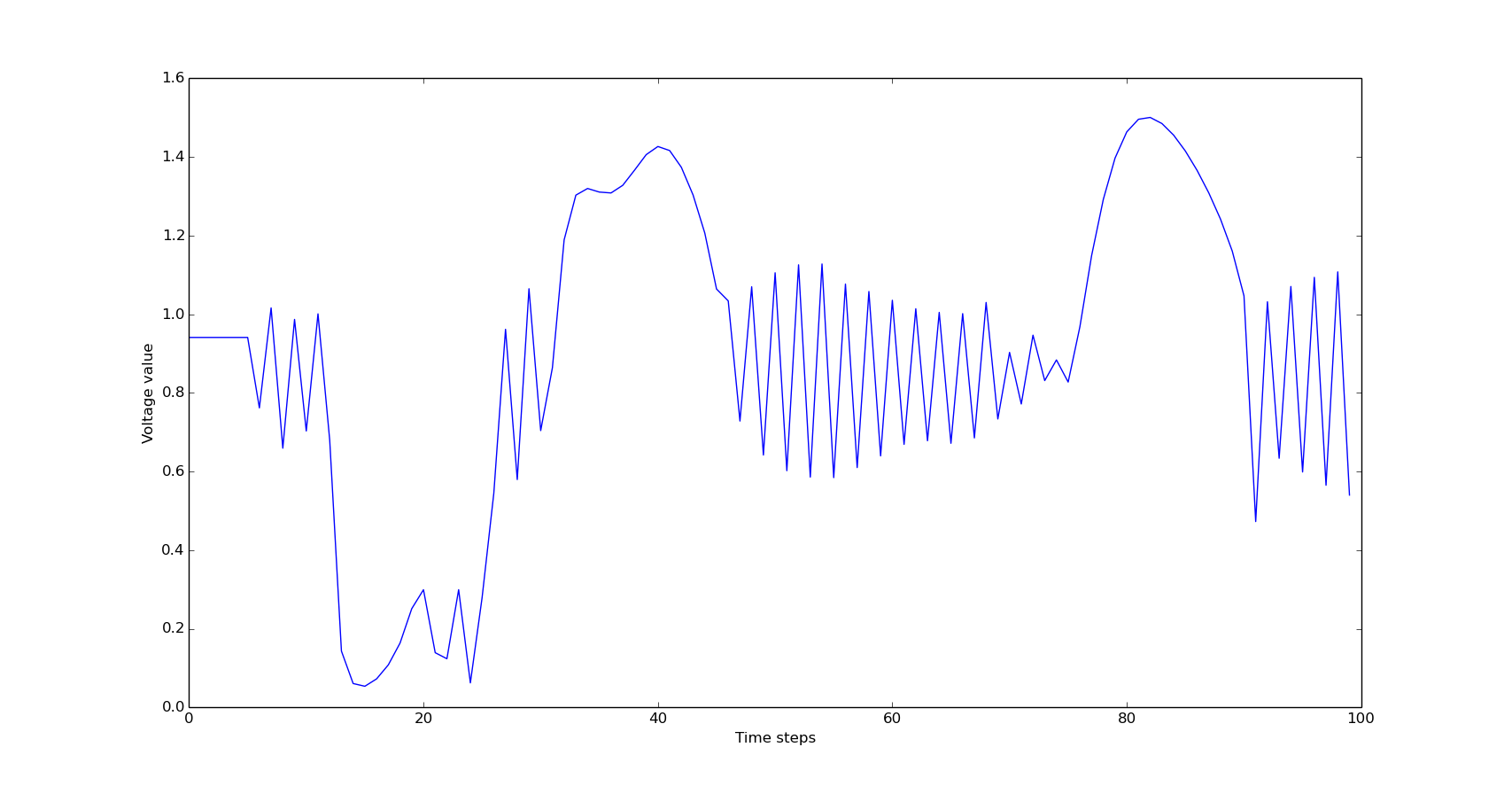}
\caption{Variation of voltage value in presence of LL fault}
\end{figure}
\begin{figure}[h]
\centering
\includegraphics[width=\linewidth, keepaspectratio]{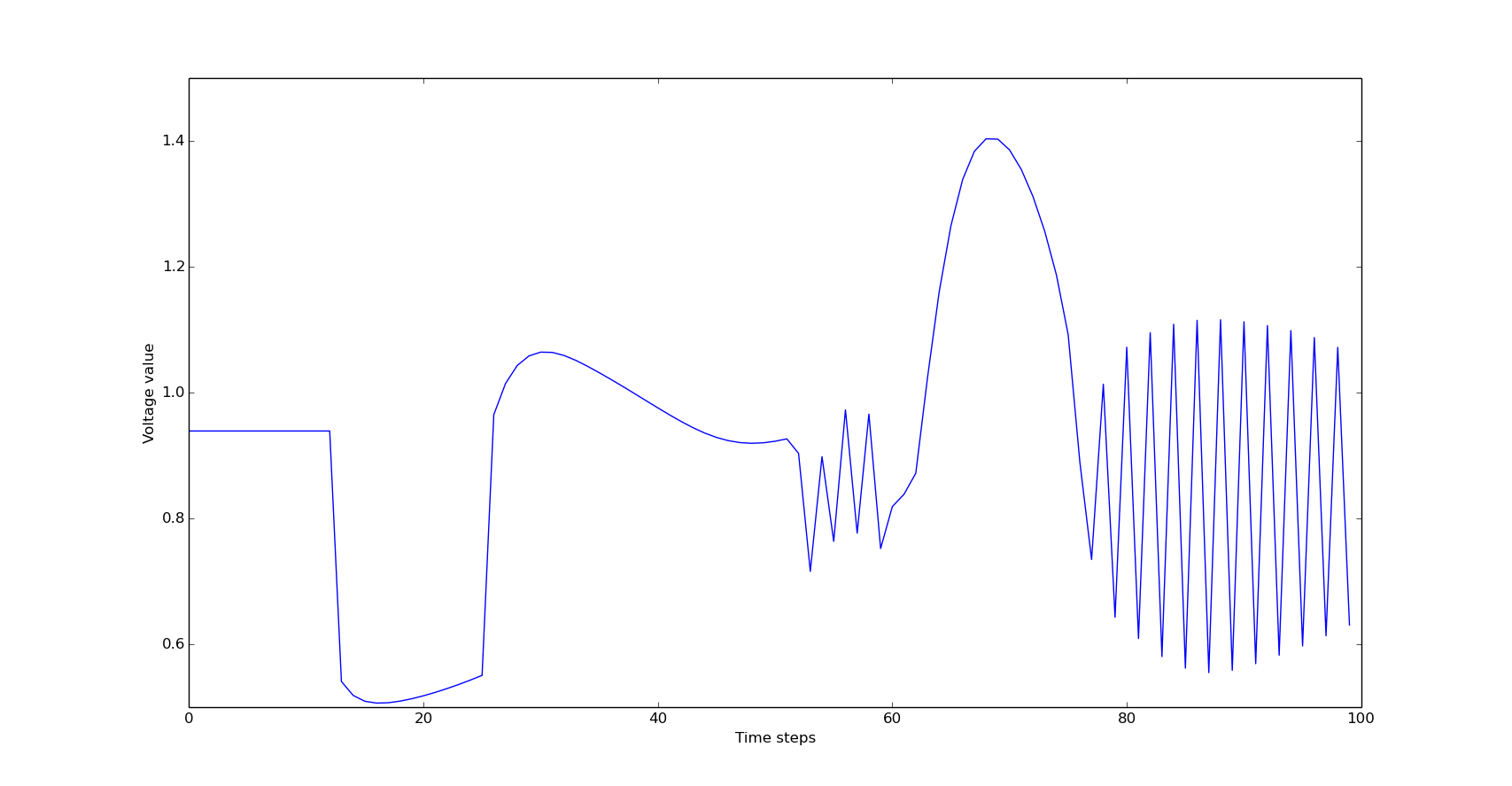}
\caption{Variation of voltage value in presence of LG fault}
\end{figure}

\subsection{Classification using SVM}
Out of the $2300$ data examples available per fault, $2000$ were used for training while the rest $300$ were used for testing. The input to the SVM model was the voltage \& angle data of all the buses for $100$ time steps. The output of the SVM classifier was the fault type or class.

The \textit{classification accuracy} on the test set was observed to be around 87-88\% for SVM classifier.

\subsection{Classification using LSTM (Recurrent neural network)}
The above SVM model had a major disadvantage in the sense that it did not utilize the information present in the data. It is the variation of voltage with time that tells us as to what fault had occurred in the network. However the data was presented just as a normal vector to the model. To utilize this time varying information models need to be constructed which take data in accordance with the variation of data with the time and this is where recurrent neural networks (RNN) come into play.

\subsubsection{The Model}
The model consists of $100$ unfoldings in time of LSTM cells corresponding to 100 time varying voltage values for each of the bus. Each LSTM cell gets a vector of size equal to the number of buses corresponding to the voltage values of the buses at that time step. Both the hidden vector and output vectors are of size $128$. The output vector of the last cell contains the temporal information present in the entire data. This information extracted can then be further used to classify the type of fault.

For the classification of faults from the information obtained from LSTM a second model is built entirely of fully connected layers. The model consists of $1$ hidden layer consisting of $64$ neurons followed by an output layer of size $2$. The output obtained determines the probability of the fault belonging to LL or LG class. The model has been described by a block diagram (Figure 8).

\begin{figure}[h]
\centering
\includegraphics[width=\linewidth, keepaspectratio]{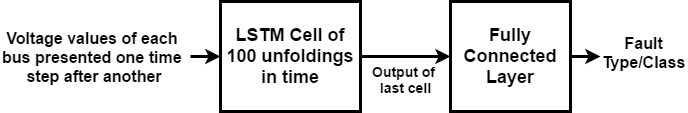}
\caption{Model using LSTM for classification of faults}
\label{fig:subim4}
\end{figure}

\subsection{Results}
With LSTM the classification accuracy jumped to 94-95\%, an \textit{improvement} of around 6\% from the SVM model. The plot of accuracy and loss with training have been shown in Figures 9 and 10.

\begin{figure}[h]
\centering
\includegraphics[width=\linewidth, keepaspectratio]{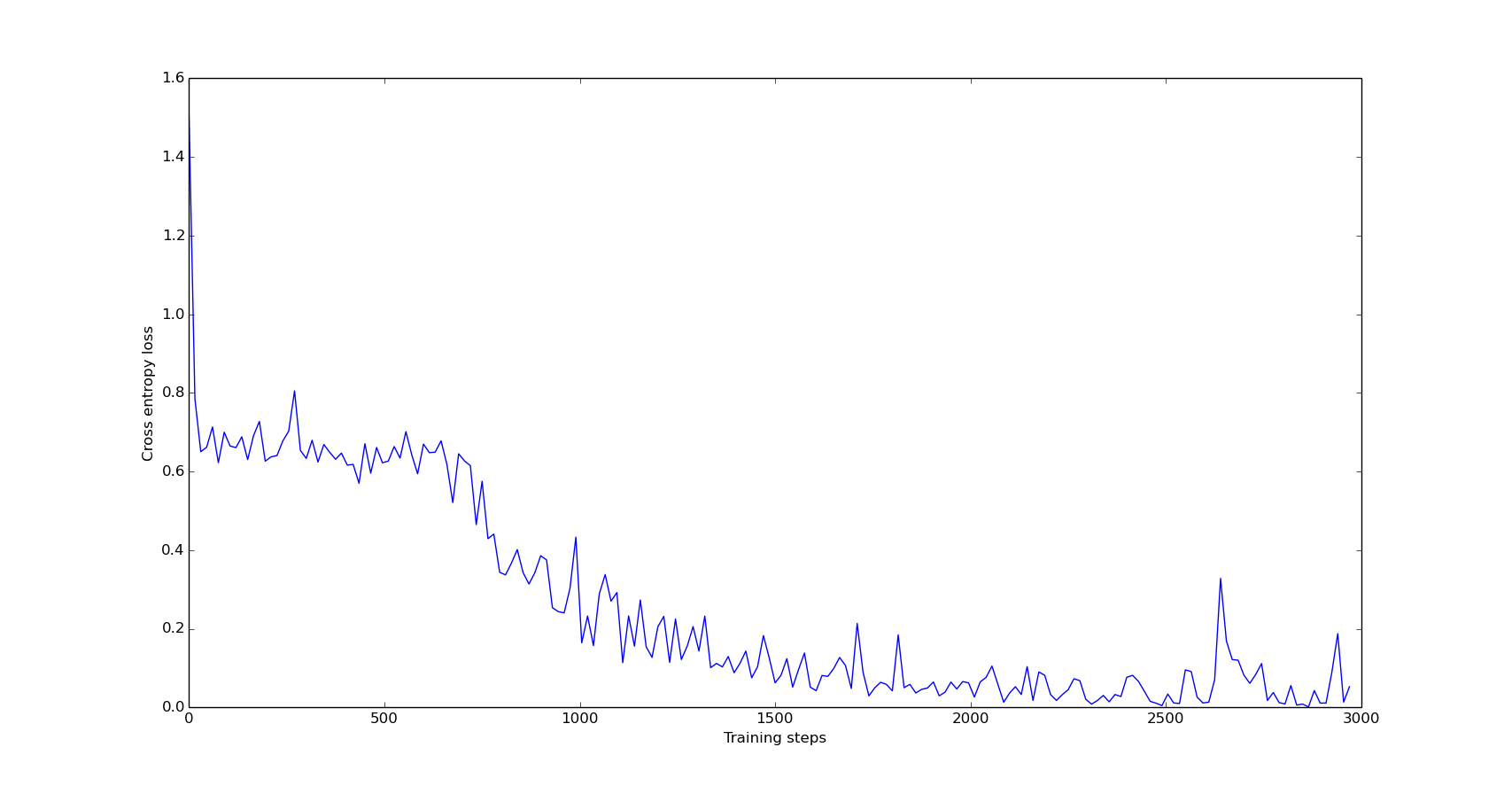}
\caption{Variation of cross entropy loss with training}
\end{figure}
\begin{figure}[h]
\centering
\includegraphics[width=\linewidth, keepaspectratio]{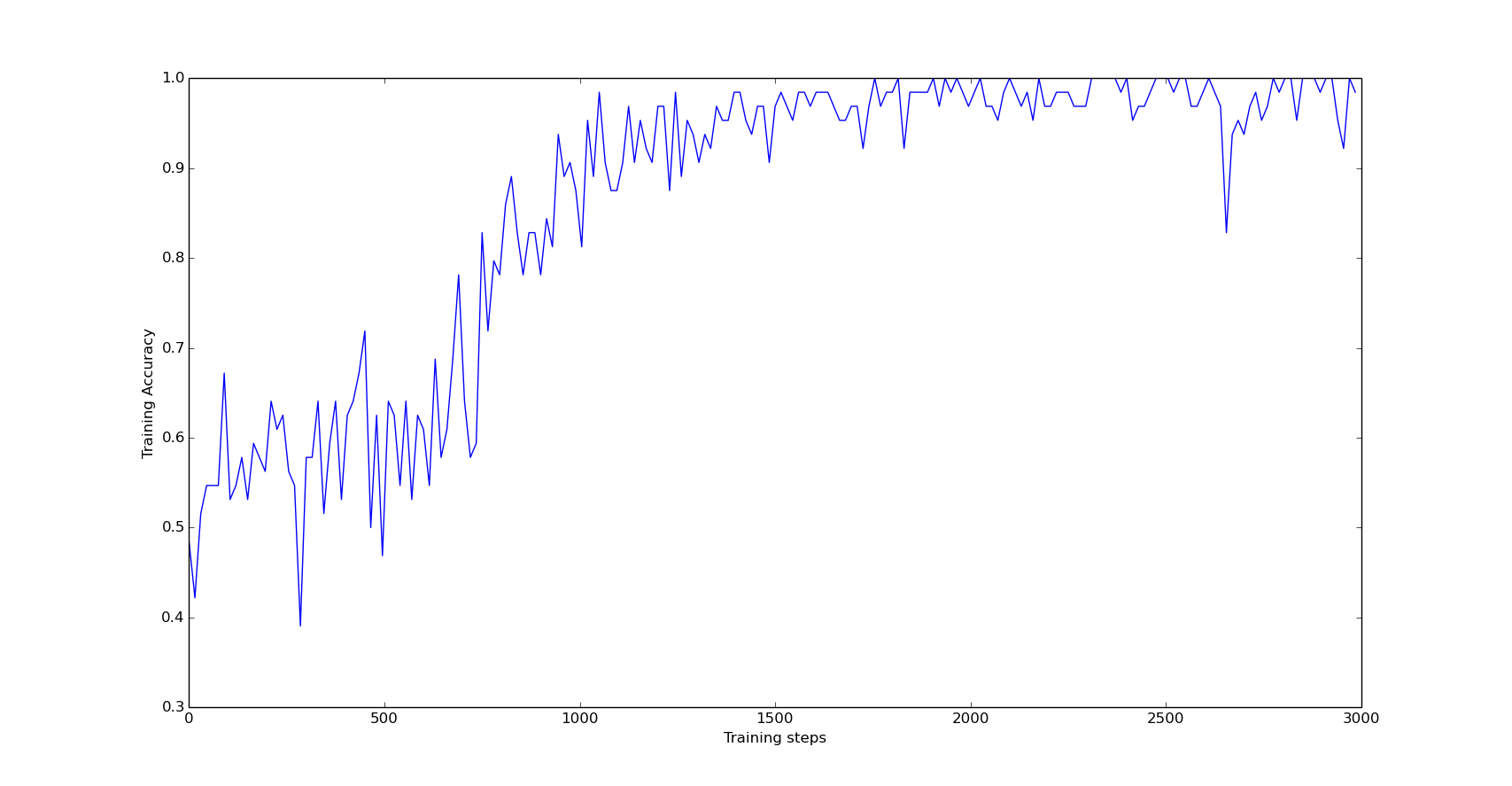}
\caption{Variation of training accuracy with progress of training}
\end{figure}

\section{Location of bus line in which fault was triggered}
Once the type of fault had been determined by the classification model, different models were constructed for each of the different fault types to determine the bus line in which the fault had been triggered. Again the voltage variations of the network with time can be used to find out the bus number. This classification task was done for both 3-phase ($3\phi$) and LL faults. The source of the fault (if any) was not part of the input for obvious reasons. This model is essential as often it is necessary to find out the location of the fault given just the PMU voltage graphs data.

\subsection{The Model}
The data input to the classifier was again the voltage data corresponding to each of the bus lines for the $100$ different steps. The classifier was trained to output the bus number corresponding to where the fault had been triggered or an output of $0$ if the network data corresponded to non-faulty one whereby no fault had been triggered on any bus line. LSTM was again used to extract meaningful time dependent information from the data which was then used to correctly classify the data via fully connected neural network layers. The fully connected layer consisted of one hidden layer consisting of $128$ neurons. The output size was one more than the number of bus lines present in the network.

The plots of the bus line in which the fault had been triggered and the other bus line where the fault had not been triggered have been shown in Figure 11.
\begin{figure}[h]
\centering
\includegraphics[width=\linewidth, keepaspectratio]{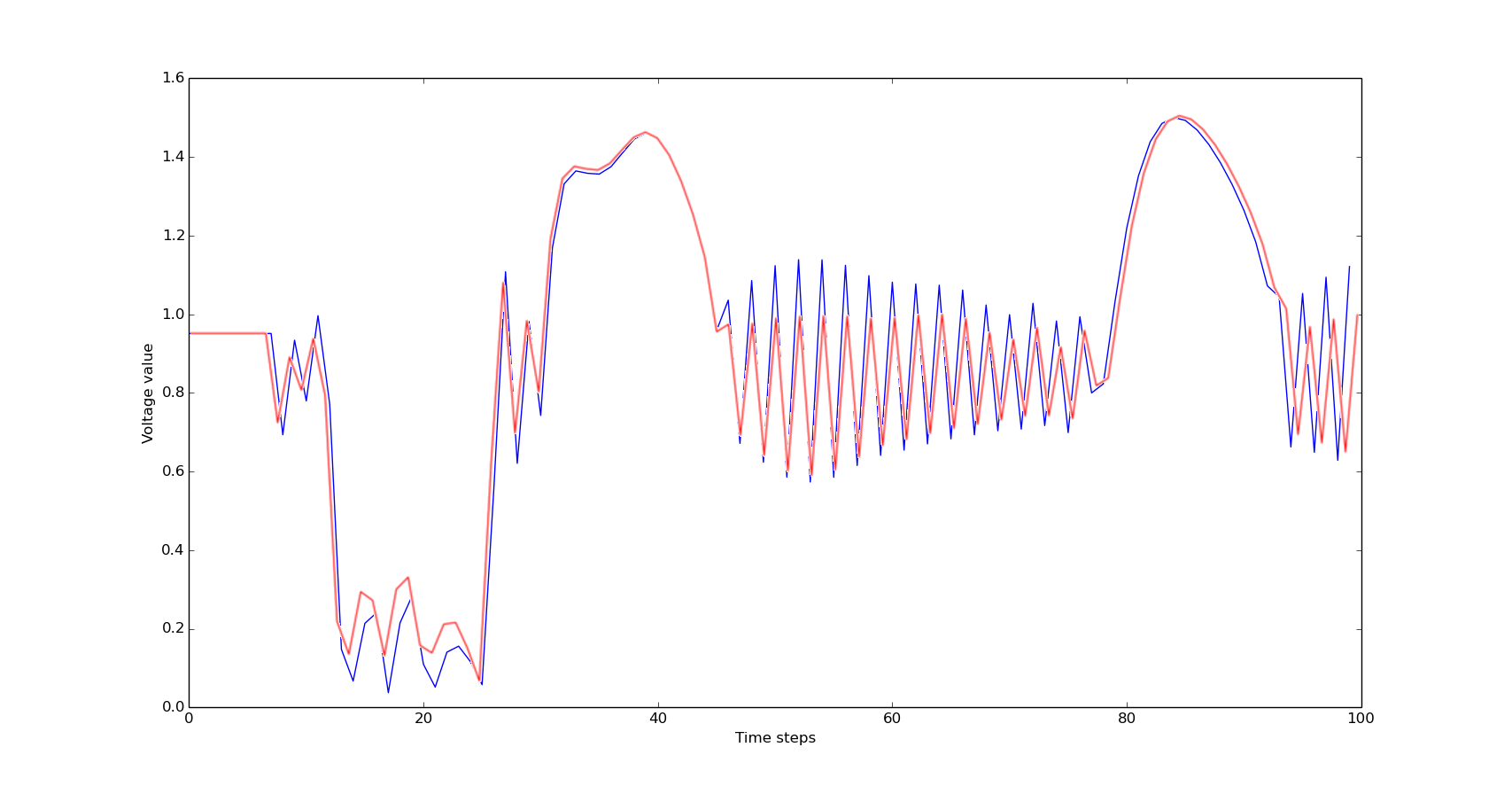}
\caption{Blue line: Voltage variation with time for the bus line in which fault was triggered; Red line: Voltage variation with time for the bus line in which no fault was triggered}
\end{figure}

\subsection{Results}
For the $3\phi$ fault, the classifier gave an accuracy of 97\% corresponding to predicting the bus number. The bus number was outputted or classified as ``$0$'' in case of non-faulty data. For the LL fault the accuracy was also 97\%. The plots of accuracy and loss with progress of training for LL fault have been shown in Figures 12 and 13.

\begin{figure}[h]
\centering
\includegraphics[width=\linewidth, keepaspectratio]{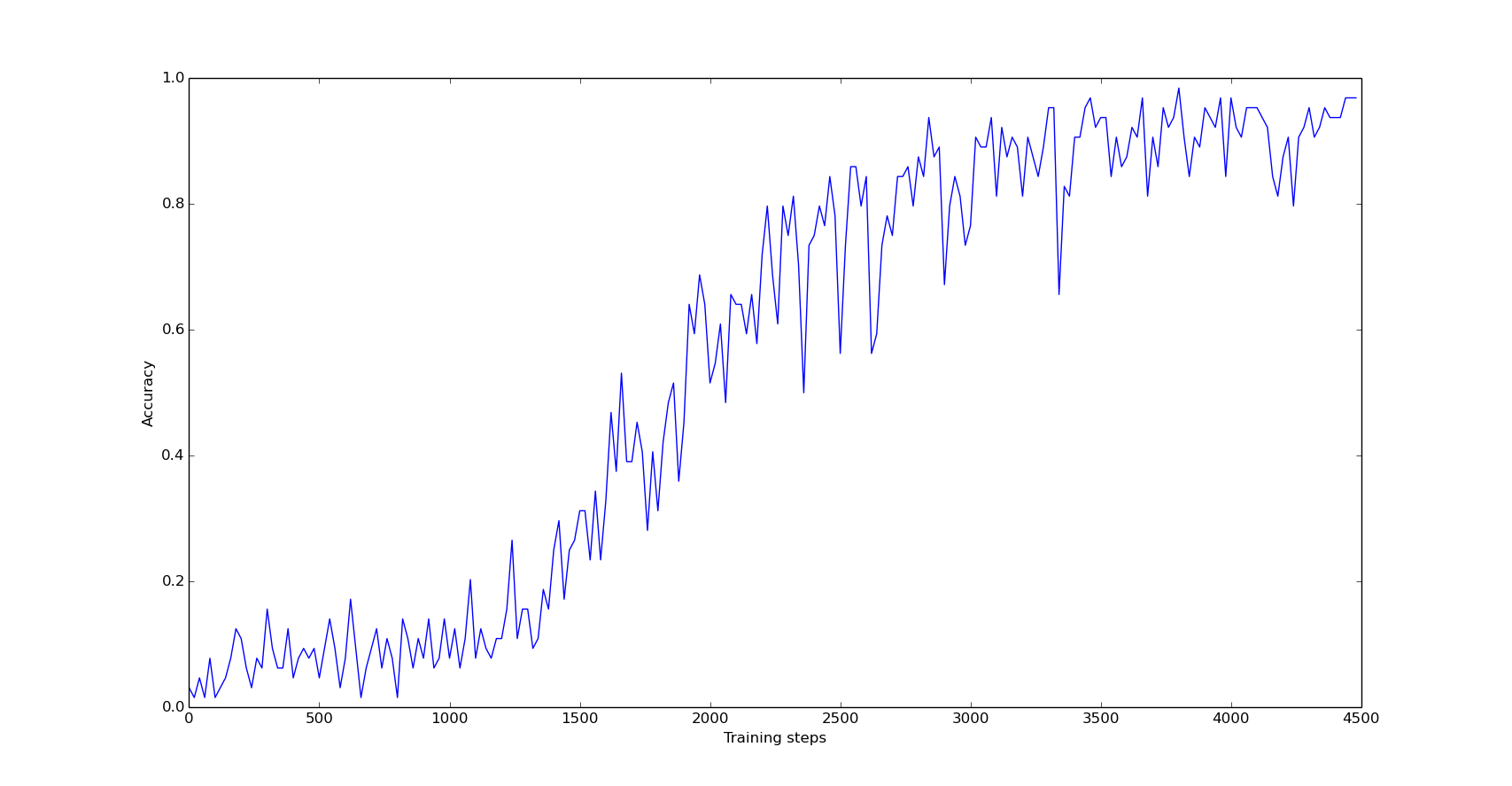}
\caption{Variation of training accuracy with progress of training}
\end{figure}

\begin{figure}[h]
\centering
\includegraphics[width=\linewidth, keepaspectratio]{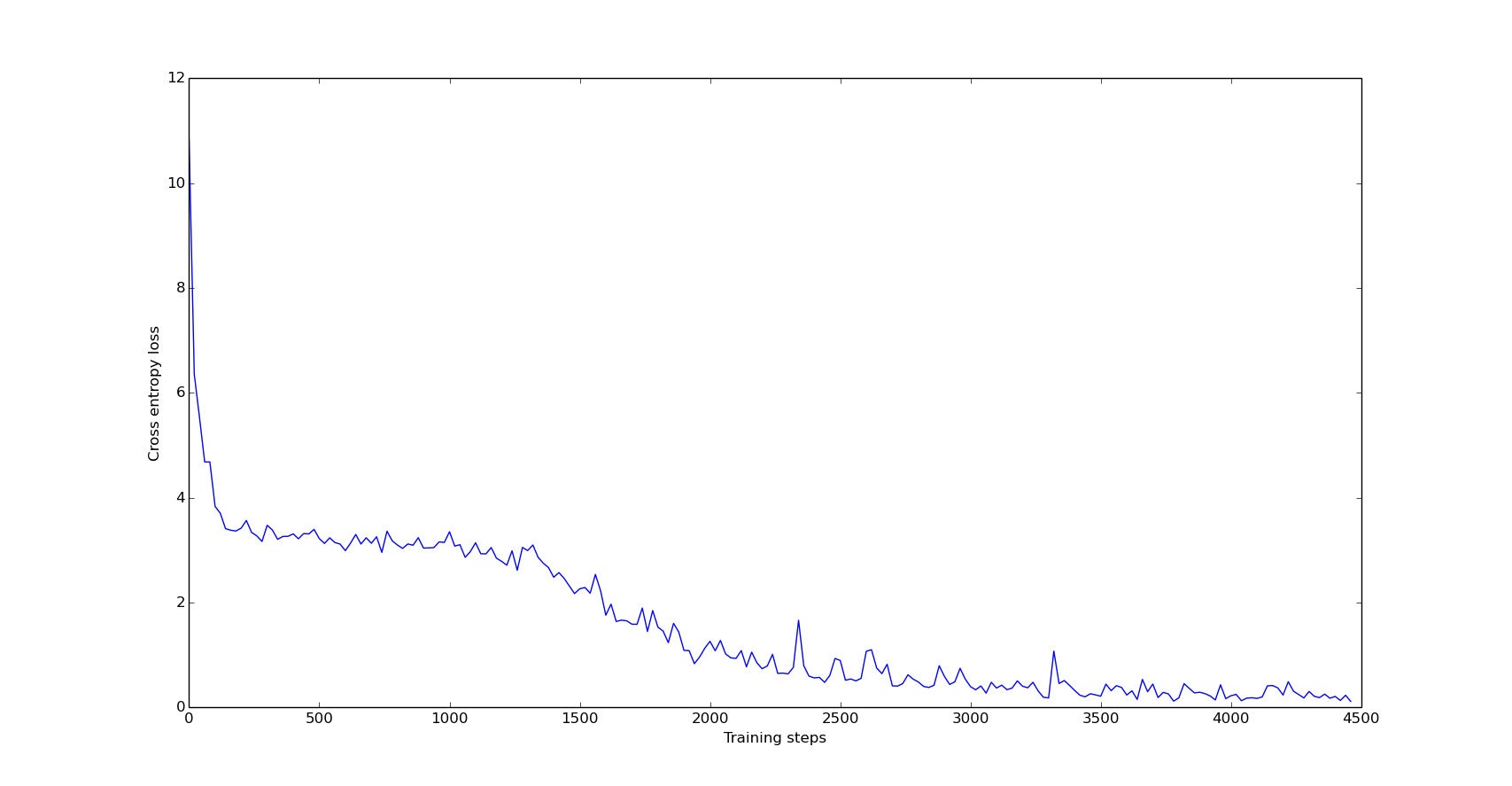}
\caption{Variation of training loss with progress of training}
\end{figure}

\section{Further Work \& Conclusion}
In conclusion, we created a grid to perform intelligent fault analysis where we forecasted the maximum voltage deviation using pre-fault data, classified the type of faults and found out the location of the fault, all using machine learning and deep learning techniques. The different models are inter-related as given the PMU data in crisis situations, running the above models will give a great deal of insight into the severity and extent of a possible undesirable situation. The results obtained were very encouraging and thus, further analysis can be done on various avenues starting from this. Some of these avenues are discussed henceforth.

Dynamic thresholds are required mainly for taking into account the difference in loading conditions. These can be developed. Another major task in the monitoring of the health of the grid is prognosis of the grid vulnerability when the actual load-generation of the grid deviates from the predicted schedule. The monitoring of grid states need to be done continuously and once the current state is found to differ significantly than the predicted state, then a close monitoring is required. The challenge is to come up with some health metrics which can, from the close monitoring of grid states, determine whether the network is going to a vulnerable state or not. The input to be considered is a time window of past network states, the current network state and the predicted state for the current time and the output will be either a yes or a no depending on whether a vulnerable state is going to be reached or not. Once the health metrics have been defined and vulnerability of the states determined, the data can then be used as training data for building or learning rules so that any new data which is not on the training set can be easily classified regarding the grid vulnerability. Issues such as congestion of the power network and optimal generator subset selection can also be explored \cite{bhattacharya1}. Finally the whole system can be tested on the real-world grid data obtained from national agencies. 

We started with an aim of deriving accurate prognostics information so as to make power grids artificially intelligent. The working system will be especially useful in renewable energy power grids where the generation periods are erratic (e.g. solar power generators generate power only during the day when the Sun is up). With the knowledge about the vulnerability of the grid, issues like load shedding, power surges etc. can be handled efficiently. With time, the system will have the ability to be more sophisticated to handle various types of networks and situations which will be suitable for deployment at the national level.

\section*{Acknowledgment}
The authors were undergraduate students of the Indian Institute of Technology (IIT) Kharagpur, West Bengal, India during the period of research for this project which culminated in this paper. The authors would like to thank the institute for their support.

\bibliographystyle{IEEEtran}
\bibliography{biblio}
\end{document}